\documentclass[preprint,12pt]{elsarticle} 
\usepackage{graphicx} 
\usepackage{amssymb} 

\journal{Nuclear Physics A} 

\begin{document}

\begin{frontmatter}

\title{Recent studies of kaonic atoms and nuclear clusters} 

\author{Avraham Gal} 
\address{Racah Institute of Physics, The Hebrew University, Jerusalem, Israel} 

\date{\today} 

\begin{abstract} 

Recent studies of kaonic atoms, few-body kaonic quasibound states and kaonic 
nuclei are reviewed, with emphasis on implementing the subthreshold energy 
dependence of the $\bar KN$ interaction in chiral interaction models that 
are consistent with the SIDDHARTA $K^-$ hydrogen data. Remarks are made on 
the possible role of the $p$-wave $\Sigma(1385)$ resonance with respect to 
that of the $s$-wave $\Lambda(1405)$ resonance in searches for strangeness 
${\cal S}=-1$ dibaryons. 

\end{abstract} 


\end{frontmatter}

\section{Introduction}
\label{sec:intro} 

Recent NLO chiral model calculations of near-threshold $\bar K N$ dynamics, 
reproducing the SIDDHARTA measurement of atomic $K^-$ hydrogen $1s$ level 
shift and width \cite{SID11}, have been discussed by Hyodo \cite{hyodo12}. 
The $\Lambda(1405)$-induced strong energy dependence of the scattering 
amplitudes $f_{\bar KN}(\sqrt{s})$ arising in these calculations introduces 
a new feature into the analysis of $K^-$ atomic and nuclear systems 
as realized for $K^-$ atoms in the early 1970s \cite{wycech71,BTo72}. Thus, 
in nuclear matter, approximated for $A\gg 1$ by the lab system, 
\begin{equation} 
s=(\sqrt{s_{\rm th}}-B_K-B_N)^2-({\vec p}_K+{\vec p}_N)^2 \leq s_{\rm th}\; , 
\label{eq:s}
\end{equation} 
where $\sqrt{s_{\rm th}}\equiv m_K+m_N$, $B_K$ and $B_N$ are binding energies, 
and where additional downward energy shift is generated by the momentum 
dependent term. Unlike in the free-space $\bar KN$ cm system where 
$({\vec p}_K+{\vec p}_N)_{\rm cm}=0$, this term is found to contribute 
substantially in the lab system in realistic applications. 
Therefore, a reliable model extrapolation of $\bar KN$ amplitudes into 
subthreshold energies is mandatory in $K^-$ atom and nuclear applications. 
 
Below I give a brief overview of works on kaonic quasibound systems and kaonic 
atoms where subthreshold $\bar K N$ amplitudes were used in a physically 
correct way during the last two years. It is shown how the energy dependence 
of these amplitudes, when translated into density dependence, leads to special 
patterns in kaonic systems. Finally, I focus attention to the recently 
proposed $I=3/2$, $J^{\pi}=2^+$ $\Sigma(1385)N$ dibaryon around the $\pi\Sigma 
N$ threshold \cite{GG13} and suggest how to search for it in experiments that 
look for the $I=1/2$, $J^{\pi}=0^-$ $\Lambda(1405)N$ dibaryon, better known as 
$K^-pp$. In lieu of a concluding section, conclusions are marked in boldface 
throughout this review.

\section{Few-body kaonic quasibound states} 
\label{sec:few} 

A prototype of such states is $K^-pp$ which stands for $\bar K NN$ with 
isospin $I=1/2$ and spin-parity $J^{\pi}=0^-$, dominated by $I_{NN}=1$ 
and $s$ waves. A summary of few-body calculations of this system is given 
in Table~\ref{tab:kpp} updating older versions in recent 
international conferences \cite{Weise10}. 

\begin{table}[htbp] 
\begin{center} 
\caption{Calculated $K^-pp$ binding energies $B$ \& widths $\Gamma$ (in MeV).} 
\begin{tabular}{lccccccc} 
\hline 
& \multicolumn{3}{c}{chiral, energy dependent} & \multicolumn{4}{c}
{non-chiral, static calculations }  \\ 
& var.~\cite{BGL12} & var.~\cite{DHW08} & Fad.~\cite{IKS10} & var.~\cite{YA02} 
& Fad~\cite{SGM07} & Fad~\cite{IS07} & var.~\cite{WG09} \\ 
\hline 
$B$ & 16 & 17--23 & 9--16 & 48 & 50--70 & 60--95 & 40--80 \\ 
$\Gamma$ & 41 & 40--70 & 34--46 & 61 & 90--110 & 45--80 & 40--85 \\ 
\hline 
\end{tabular} 
\label{tab:kpp} 
\end{center} 
\end{table} 

The listed calculations are $\bar KNN$ variational (var.) where the complex 
$\bar KN$ interaction accounts for the $\bar KN$--$\pi\Sigma$ two-body coupled 
channels but disregards $\bar KNN$--$\pi\Sigma N$ coupling, or fully coupled 
channels three-body Faddeev (Fad.). A more revealing classification of these 
calculations is according to whether or not the input two-body interactions 
are energy dependent. The table makes it clear that the binding energies 
calculated by using chiral, energy dependent interactions are considerably 
lower than those calculated using energy independent interactions, the reason 
for which is the marked difference between the $(\bar KN)_{I=0}$ interaction 
strengths which yield a quasibound state at $\approx$1420 MeV in the former 
case and at $\approx$1405 MeV in the latter case. 

The recent calculations by Barnea et al. \cite{BGL12} include on top of 
$K^-pp$ also the four-body $\bar KNNN$ quasibound states with $I=0,1$ and 
the $I=0$ lowest $\bar K\bar KNN$ quasibound state. The calculations were 
done extending a nuclear hyperspherical basis to include $\bar K$ mesons. 
The $A$-body wavefunctions were expanded in this complete basis and (real) 
ground-state binding energies were computed variationally. Convergence was 
assessed by increasing systematically the size of the basis used. For input 
two-body interactions, the AV4' $V_{NN}$ was used together with an effective 
energy-dependent complex $V_{\bar KN}$ \cite{HW08} and a weakly repulsive 
$V_{\bar K\bar K}$ \cite{KEJ08}. In single-$\bar K$ configurations, 
$V_{\bar KN}$ was evaluated at subthreshold energies obtained by expanding 
Eq.~(\ref{eq:s}) nonrelativistically near $\sqrt{s_{\rm th}}$: 
\begin{equation} 
\sqrt{s} = \sqrt{s_{\rm th}}-\frac{B}{A}-\frac{A-1}{A}B_K-
\xi_{N}\frac{A-1}{A}\langle T_{N:N} \rangle -\xi_{K}\left ( \frac{A-1}{A} 
\right )^2 \langle T_K \rangle \; ,
\label{eq:sqrt{s}} 
\end{equation} 
where $\xi_{N(K)}\equiv m_{N(K)}/(m_N+m_K)$, $B$ is the total binding energy 
of the system and $B_K=-E_K$, $T_K$ is the kaon kinetic energy operator in 
the total cm frame and $T_{N:N}$ is the pairwise $NN$ kinetic energy operator 
in the $NN$ pair cm system. This expression provides a self-consistency cycle 
by requiring that $\sqrt{s}$ derived through Eq.~(\ref{eq:sqrt{s}}) from the 
solution of the Schroedinger equation agrees with the value of $\sqrt{s}$ used 
for the input $V_{\bar K N}(\sqrt{s})$. A similar expression was obtained for 
${\bar K}{\bar K}NN$ configurations. The $\bar K N \to \pi Y$ widths of these 
few-body systems were evaluated using the expression  
\begin{equation} 
\frac{\Gamma}{2}\approx\langle \,\Psi_{\rm g.s.} |
-{\rm Im}\,{\cal V}_{\bar KN}\, | \, \Psi_{\rm g.s.} \, \rangle \;, 
\label{eq:Gamma} 
\end{equation} 
where ${\cal V}_{\bar KN}$ sums over all pairwise $\bar KN$ interactions. 
Expression (\ref{eq:Gamma}) provides a good approximation owing to $|{\rm Im}
\,{\cal V}_{\bar KN}|\ll |{\rm Re}\,{\cal V}_{\bar KN}|$ \cite{HW08}. 
 
\begin{figure}[htbp] 
\begin{center} 
\includegraphics[width=0.8\columnwidth]{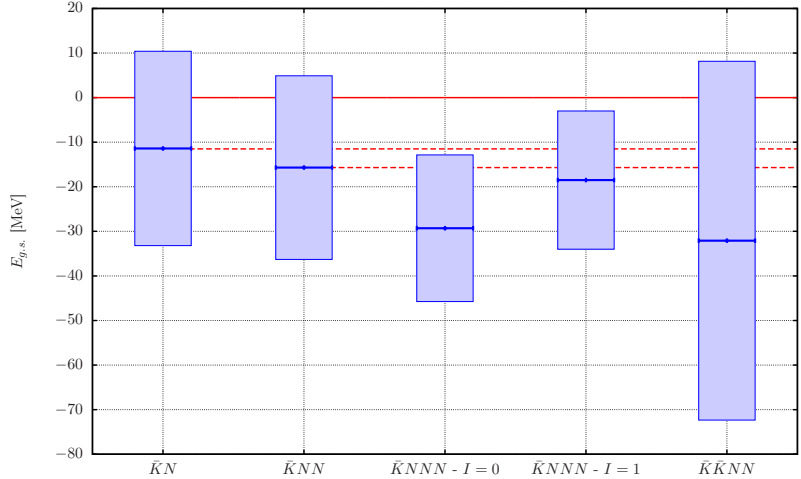} 
\end{center} 
\caption{Calculated binding energies and $\bar K N \to \pi Y$ widths of 
$\bar K$ and $\bar K\bar K$ few-body quasibound states \cite{BGL12} in MeV. 
Horizontal lines denote particle-stability thresholds and widths are 
represented by vertical bars.} 
\label{fig:gal1} 
\end{figure} 

Results of self consistent calculations are shown in Fig.~\ref{fig:gal1} 
with relevant thresholds marked in red horizontal lines. Since both 
$|{\rm Re}\,V_{\bar KN}|$ and $|{\rm Im}\,V_{\bar KN}|$ decrease upon 
going subthreshold \cite{HW08}, the self-consistently calculated binding 
energies (widths) come out typically 10 (10--40) MeV lower than when 
$V_{\bar KN}(\sqrt{s_{\rm th}})$ is imposed. The $I=1/2$ $\bar KNN$ g.s. 
corresponds to the most left-hand side `$K^-pp$' calculation listed in 
Table~\ref{tab:kpp}. It lies only 4.3 MeV below the 11.4 MeV centroid of the 
$I=0$ $\bar KN$ quasibound state, the latter value differing substantially 
from the value 27 MeV borrowed from the $\Lambda(1405)$ resonance in 
non-chiral calculations. The next two quasibound states shown in the figure 
are the lowest $I=0$ and $I=1$ $\bar KNNN$ quasibound states, the $I=0$ state 
is the one predicted first by Akaishi and Yamazaki to be bound by over 100 MeV 
\cite{AY02}. Finally, the calculated 32 MeV binding energy of $\bar K\bar KNN$ 
is substantially lower than the one advocated for it by Yamazaki et al. 
\cite{YDA04}. {\bf With total binding energies of order 30 MeV calculated by 
Barnea et al. for the $A=4$ kaonic clusters \cite{BGL12}, there is little 
incentive to argue that $\bar K$ mesons offer a realization of strange 
hadronic matter in nature, more than $\Lambda$ and $\Xi$ hyperons do 
\cite{GFGM09}.} The widths exhibited in Fig.~\ref{fig:gal1}, of order 
40 MeV for single-$\bar K$ clusters and twice as much for double-$\bar K$ 
clusters, are due to $\bar K N \to \pi Y$. Two-nucleon absorption widths 
$\Delta\Gamma_{\rm abs}$ accounting for the poorly understood non-pionic 
processes $\bar K NN\to YN$ add $\Delta\Gamma_{\rm abs}(K^-pp)\lesssim 10$~MeV 
in $K^-pp$ \cite{DHW08} and $\sim$20 MeV in the 4-body systems \cite{BGL12}. 
{\bf Given the low binding energies and sizable widths in chirally motivated 
calculations, a clear identification of such near-threshold quasibound states 
in ongoing experimental searches is likely to be difficult.} For $K^-pp$, 
in particular, a ${\bar K}^0d$ quasibound state contamination cannot be 
ruled out, with arguments for and against its presence that may be deduced 
from Refs.~\cite{Bayar12,Shev12} respectively.

\section{Kaonic atoms} 
\label{sec:atoms} 

\begin{figure}[htbp] 
\begin{center} 
\includegraphics[width=0.48\columnwidth]{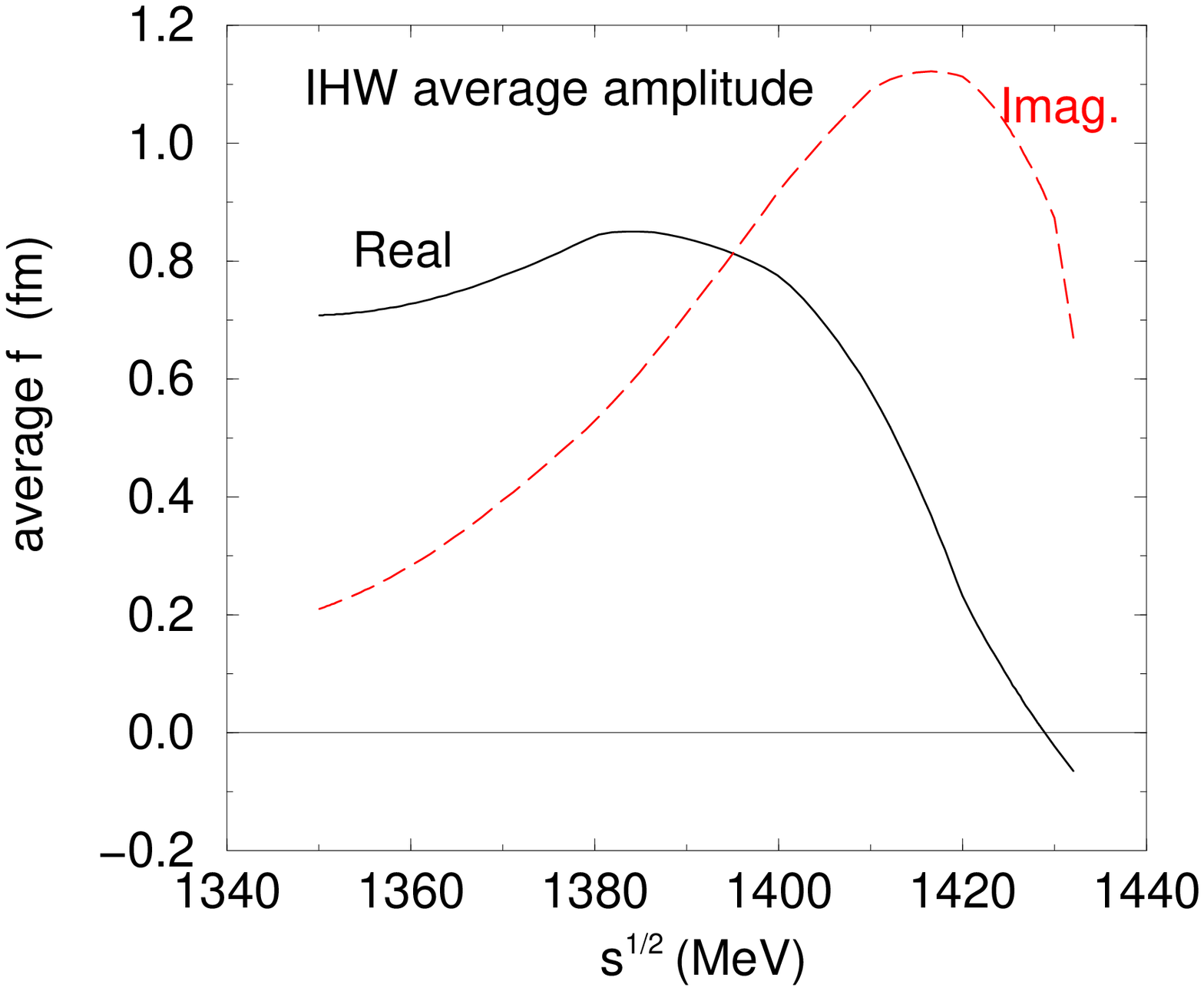} 
\includegraphics[width=0.48\columnwidth]{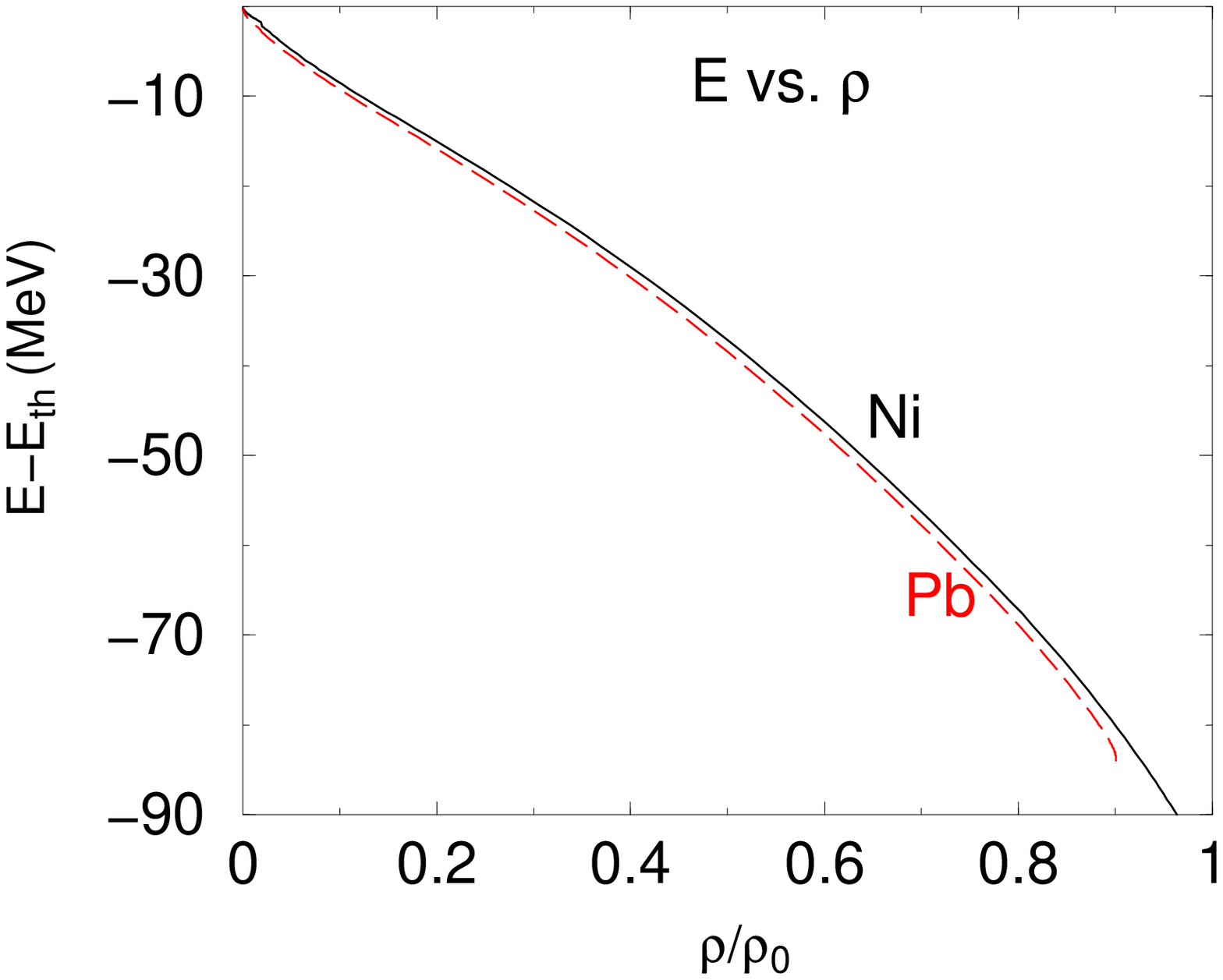} 
\end{center} 
\caption{Left: the IHW cm scattering amplitude $f_{K^-N}(\sqrt{s})=
\frac{1}{2}(f_{K^-p}(\sqrt{s})+f_{K^-n}(\sqrt{s}))$ \cite{IHW11}. 
Right: subthreshold $K^-N$ energies probed by $V_{K^-}$ as a function 
of nuclear density in Ni and Pb, calculated self consistently from the 
IHW-based best fit to kaonic atoms specified in Table~\ref{tab:DD}.} 
\label{fig:gal2} 
\end{figure}

Here I review the latest kaonic atom calculation \cite{FG13}, using for input 
the recent Ikeda-Hyodo-Weise (IHW) NLO chiral $K^-N$ subthreshold scattering 
amplitudes \cite{IHW11} which are constrained by the kaonic hydrogen SIDDHARTA 
measurement \cite{SID11}. Shown on the l.h.s. of Fig.~\ref{fig:gal2} is the 
free-space charge averaged $K^-N$ c.m. subthreshold scattering amplitude 
$f_{K^-N}(\sqrt{s})$. Its substantial energy dependence was converted to 
density dependence by the expression 
\begin{equation} 
(\sqrt{s_{\rho}})_{\rm atom}\approx\sqrt{s_{\rm th}}-B_N(\rho/{\bar\rho})
-15.1(\rho/\rho_0)^{2/3} 
 +\xi_K({\rm Re}~V_{K^-}+V_{\rm Coul} (\rho/\rho_0)^{1/3}) 
\label{eq:SC} 
\end{equation} 
(in MeV), obtained by taking the limit $A\gg 1$ in Eq.~(\ref{eq:sqrt{s}}) 
where the nuclear kinetic energy was evaluated within the Fermi gas model, 
giving rise to a ${\rho}^{2/3}$ dependence, and the $K^-$ kinetic energy 
was traded in favor of ${\rm Re}V_{K^-}$+$V_{\rm Coul}$ in the local density 
approximation. For atoms, $B_K=0$ was assumed. The density dependence 
attached to $B_N$ (where $\bar\rho$ is the {\it average} nuclear density) 
and to $V_{\rm Coul}$ ensures that the low-density limit is satisfied. 
The resulting self-consistent application of the constraint (\ref{eq:SC}) 
in kaonic atom fits is shown on the r.h.s. of Fig.~\ref{fig:gal2} where 
subthreshold $K^-N$ energies probed by the self-consistently fitted $K^-$ 
nuclear potential at threshold are plotted as a function of nuclear density 
in Ni and Pb. It is seen that the energy downward shift at 50\% of central 
nuclear density amounts to $\approx$40~MeV. 

The $K^-$ nuclear potential employed in the kaonic atom fits is composed of 
two pieces, $V^{(1)}_{K^-}(\rho)$ and $V^{(2)}_{K^-}(\rho)$. The first one, 
$V^{(1)}_{K^-}(\rho)$, is derived from an underlying free-space chiral model. 
In Refs.~\cite{CFGGM11,FG12} in-medium amplitudes were obtained directly by 
solving the in-medium coupled-channels Lippmann-Schwinger equations as a 
function of the nuclear density $\rho$, using separable interaction models 
by Ciepl\'{y} and Smejkal \cite{CS12}, see for example Fig.~1 of Gazda and 
Mare\v{s} in these proceedings \cite{GM12}. In the very recent kaonic atom 
analysis by Friedman and Gal \cite{FG13} the free-space IHW cm amplitudes $f$ 
were used as input in a multiple scattering approach \cite{WRW97} to generate 
in-medium cm amplitudes ${\cal F}_{K^-N}(\rho)$, dominated by Pauli 
correlations, viz. 
\begin{equation} 
{\cal F}_{K^-N}(\rho)= \frac{(2f_{K^-p}-f_{K^-n})\:
\frac{1}{2}\rho_p}{1+\xi(\rho){\tilde f}_{I=0}\rho(r)}+\frac{f_{K^-n}
(\frac{1}{2}\rho_p+\rho_n)}{1+\xi(\rho){\tilde f}_{I=1}\rho(r)} \; , 
\label{eq:WRW} 
\end{equation} 
where ${\tilde f}=(\sqrt{s}/m_N)f$ is a lab amplitude, 
and $\xi(\rho)=9\pi/4p_F^2$ with $p_F$ the Fermi momentum. 
$V^{(1)}_{K^-}(\rho)$ is then given by 
\begin{equation} 
2{\mu_K}V^{(1)}_{K^-}(\rho)=-\:4\pi\:{\tilde{\cal F}}_{K^-N}(\rho) \; .  
\label{eq:calF} 
\end{equation} 

\begin{figure}[htbp] 
\begin{center} 
\includegraphics[width=0.48\columnwidth]{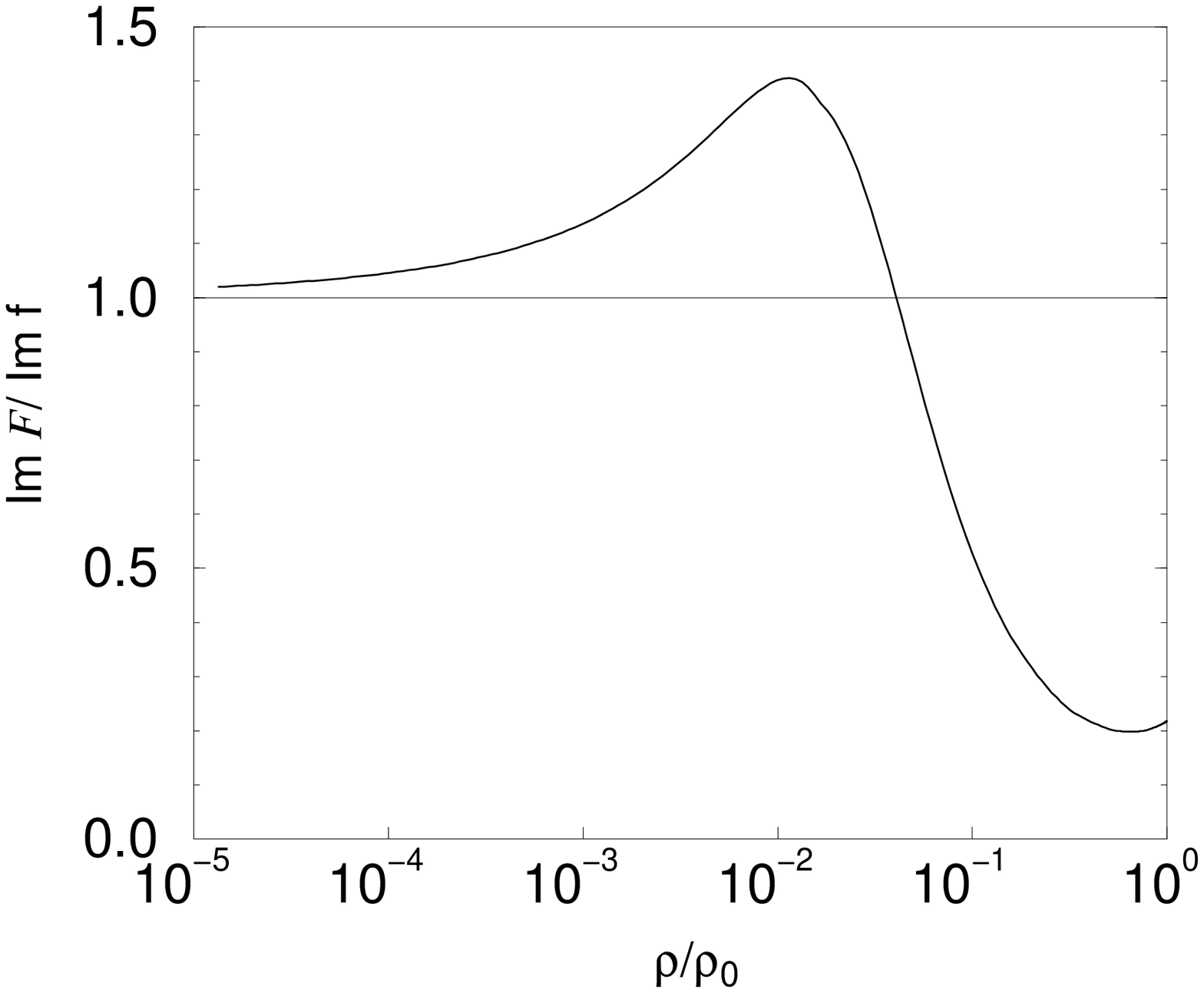} 
\includegraphics[width=0.48\columnwidth]{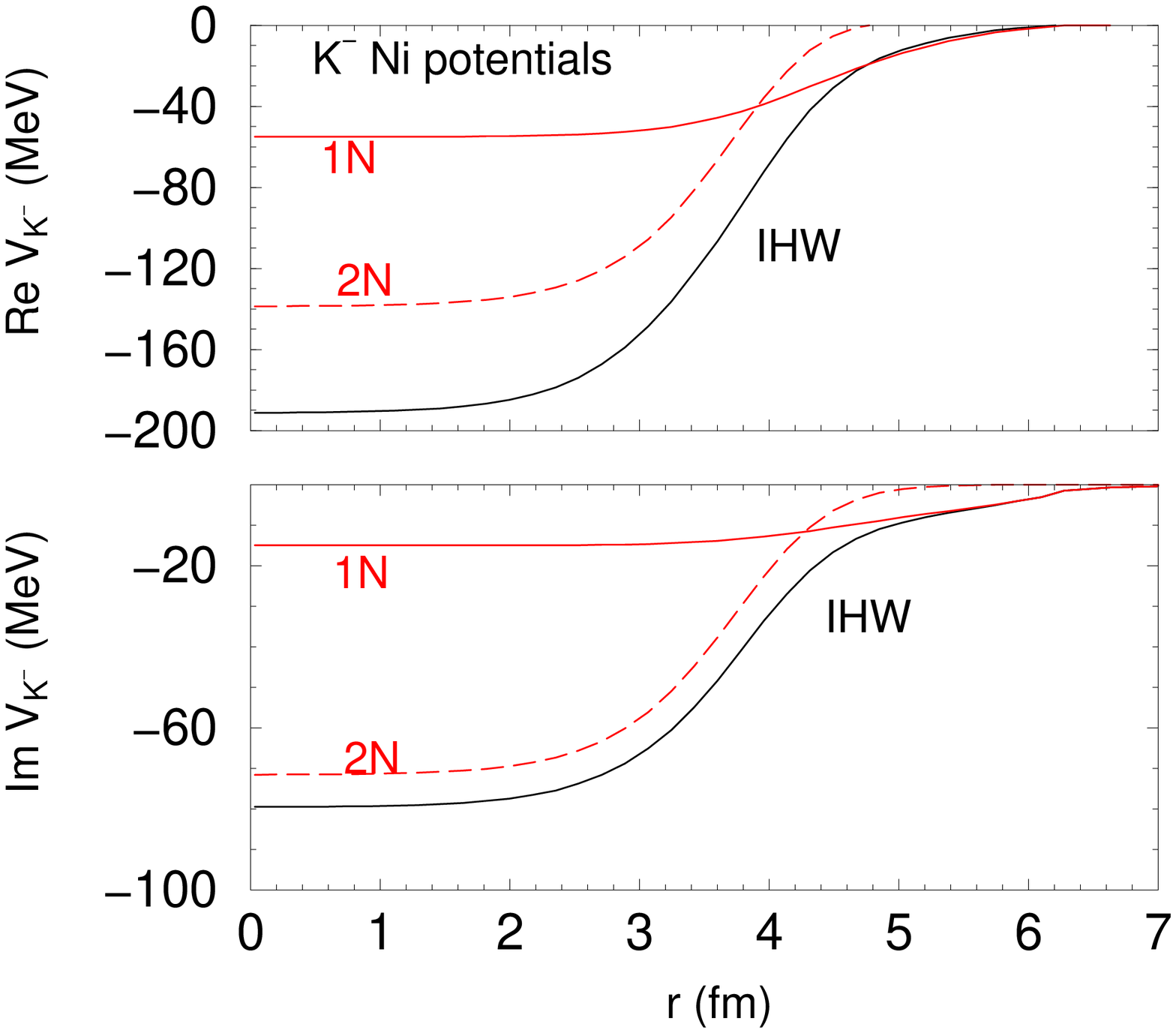} 
\end{center} 
\caption{Left: ratio of Im~${\cal F}_{K^-N}(\rho)$ to Im~$f_{K^-N}(\rho)$ 
calculated as a function of nuclear density in Ni for the IHW amplitudes 
in the absence of the $V^{(2)}_{K^-}$ potential term. Right: 1$N$ and 2$N$ 
components of the IHW-based kaonic atom potential in Ni.} 
\label{fig:gal3} 
\end{figure}

Since absorptivities turn out to play a key role in the analysis of kaonic 
atoms, I demonstrate on the l.h.s. of Fig.~\ref{fig:gal3} the ratio between 
the imaginary parts of ${\cal F}_{K^-N}(\rho)$ and $f_{K^-N}(\rho)$, plotted 
on a logarithmic density scale to highlight its slow convergence to the low 
density limit of 1. The slow convergence is caused by the predominance of the 
$\Lambda(1405)$ resonance for densities roughly below 0.06$\rho_0$ where 
this ratio exhibits hump structure with values exceeding 1, owing to the 
large negative values assumed by Re~${\tilde f}_{I=0}$ near threshold in 
Eq.~(\ref{eq:WRW}). At densities above 0.06$\rho_0$, the shown ratio 
decreases monotonically with density from a value 1 owing to the rapid 
increase of Re~${\tilde f}_{I=0}$ below 1415~MeV (where the $\Lambda(1405)$ 
resonates as far as $f$ is concerned) and levels off when 
Re~${\tilde f}_{I=0}$ has reached its (positive) maximum value. Thus, although 
the effect of the $\Lambda(1405)$ subthreshold resonance appears to be limited 
to low densities, it affects implicitly the absorptivity Im~${\cal F}_{K^-N}
(\rho)$ at densities higher than $\approx$0.1$\rho_0$, cutting it to less than 
half of the free-space input absorptivity Im~$f_{K^-N}(\rho)$. The resulting 
Im~${\cal F}_{K^-N}(\rho)$ which is induced by $K^-N\to\pi Y$ pionic 
absorption is totally insufficient in kaonic atoms fits and needs to be 
supplemented by non-pionic absorptive contributions. 

The second piece of the $K^-$ nuclear potential is a phenomenological 
$V^{(2)}_{K^-}$, simulating two-nucleon (2$N$) dispersive and absorptive 
$K^-NN \to YN$ processes and multi-nucleon processes. It is parametrized by 
\begin{equation} 
V^{(2)}_{K^-}=-(4\pi/2\mu_K )[b\rho+B\rho(\rho/\rho_0)^\alpha] \; , 
\label{eq:phen} 
\end{equation} 
where $\mu_K$ is the kaon-nucleus reduced mass, $b$ and $B$ are two complex 
parameters and the real exponent $\alpha$ satisfies $\alpha > 0$. Constraining 
Im~$V^{(2)}_{K^-}$ to be non-negative at all values of $\rho$ by imposing 
Im~$b$=0 and griding on $\alpha$, the following values are obtained in 
a best-fit search: 
\begin{equation} 
b=(-0.34\pm 0.07)~{\rm fm}, \;\; 
B=(1.94\pm 0.16)+{\rm i}(0.83\pm 0.16)~{\rm fm}, \;\; \alpha=1.2 \;. 
\label{eq:parameters} 
\end{equation} 
{\bf The contribution of this fitted $V^{(2)}_{K^-}$ to the kaonic atoms 
potential is substantial for densities larger than $\approx$0.5$\rho_0$ where 
it exceeds the contribution of the IHW-based $V^{(1)}_{K^-}$, as clearly seen 
on the r.h.s. of Fig.~\ref{fig:gal3}.} The plotted 1$N$ potential corresponds 
to a situation where $V^{(2)}_{K^-}=0$, but in the final 1$N$+2$N$ potential 
the 1$N$ component is affected by the 2$N$ component due to the implicit 
coupling between the two in the self consistent procedure related to 
Eq.~(\ref{eq:SC}). A non-additivity of the two components in the final 
potential is observed, particularly for the imaginary potential. 

\begin{table}[htbp] 
\begin{center} 
\caption{Properties of $K^-$--Ni potentials from global fits to 65 kaonic 
atom data points. Values of potentials are in MeV, r.m.s. radii in fm.} 
\begin{tabular}{lcccccc} 
\hline 
model & $\chi^2$ & $V_{\rm R}(0)$ & $V_{\rm I}(0)$ & 
$r_{\rm R}$ & $r_{\rm I}$ & $\alpha$ \\ \hline 
DD \cite{FGB94}   & 103 & $-$ 199 & $-$76 & 3.48 & 3.71 & 0.25 \\ 
IHW \cite{FG13}   & 118 & $-$ 191 & $-$79 & 3.34 & 3.73 & 1.2  \\  
NLO30 \cite{FG12} & 148 & $-$ 179 & $-$71 & 3.42 & 3.70 & 1.0  \\
\hline 
\end{tabular} 
\label{tab:DD} 
\end{center} 
\end{table} 

\begin{figure}[htbp] 
\begin{center} 
\includegraphics[width=0.48\columnwidth]{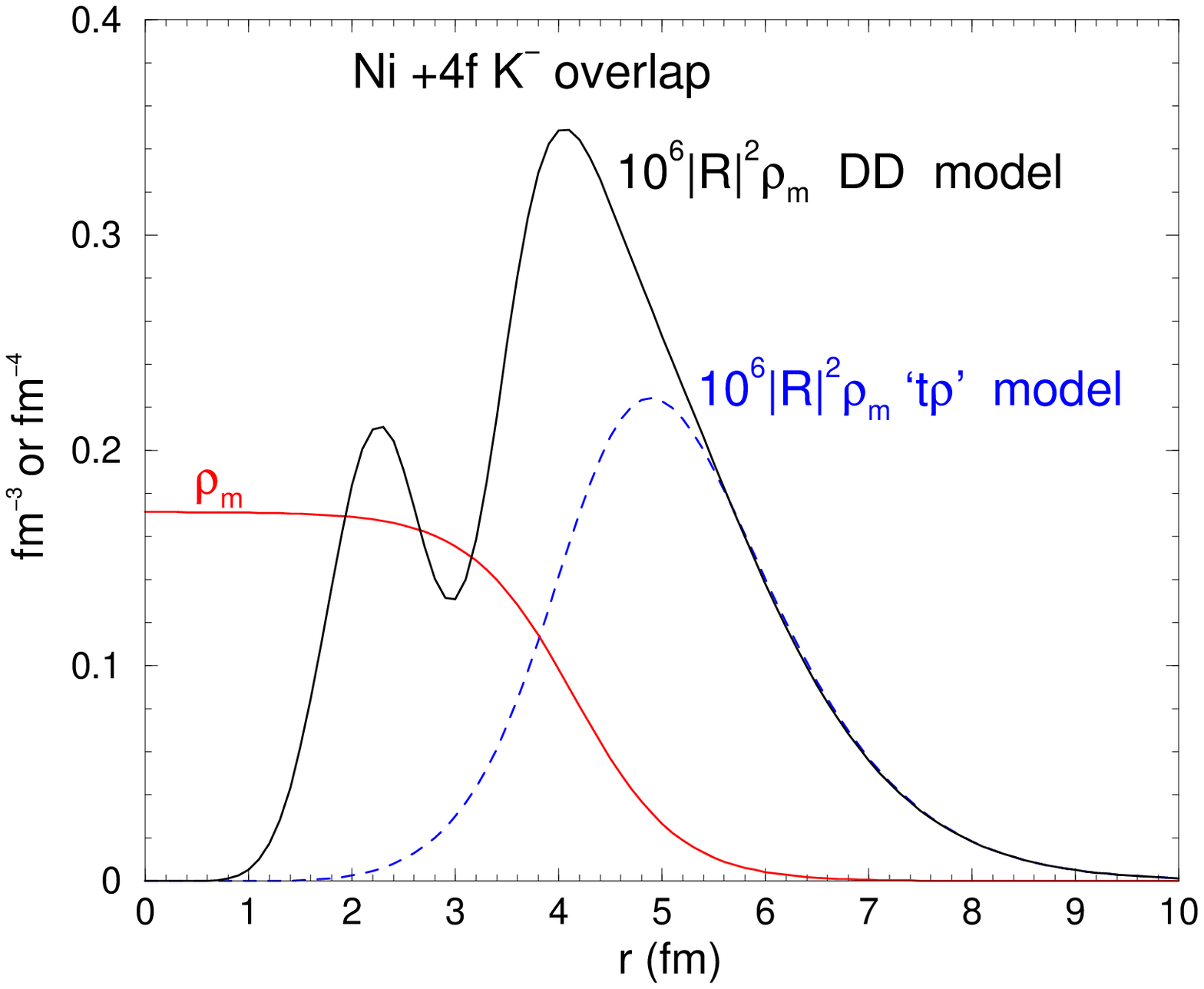} 
\includegraphics[width=0.48\columnwidth]{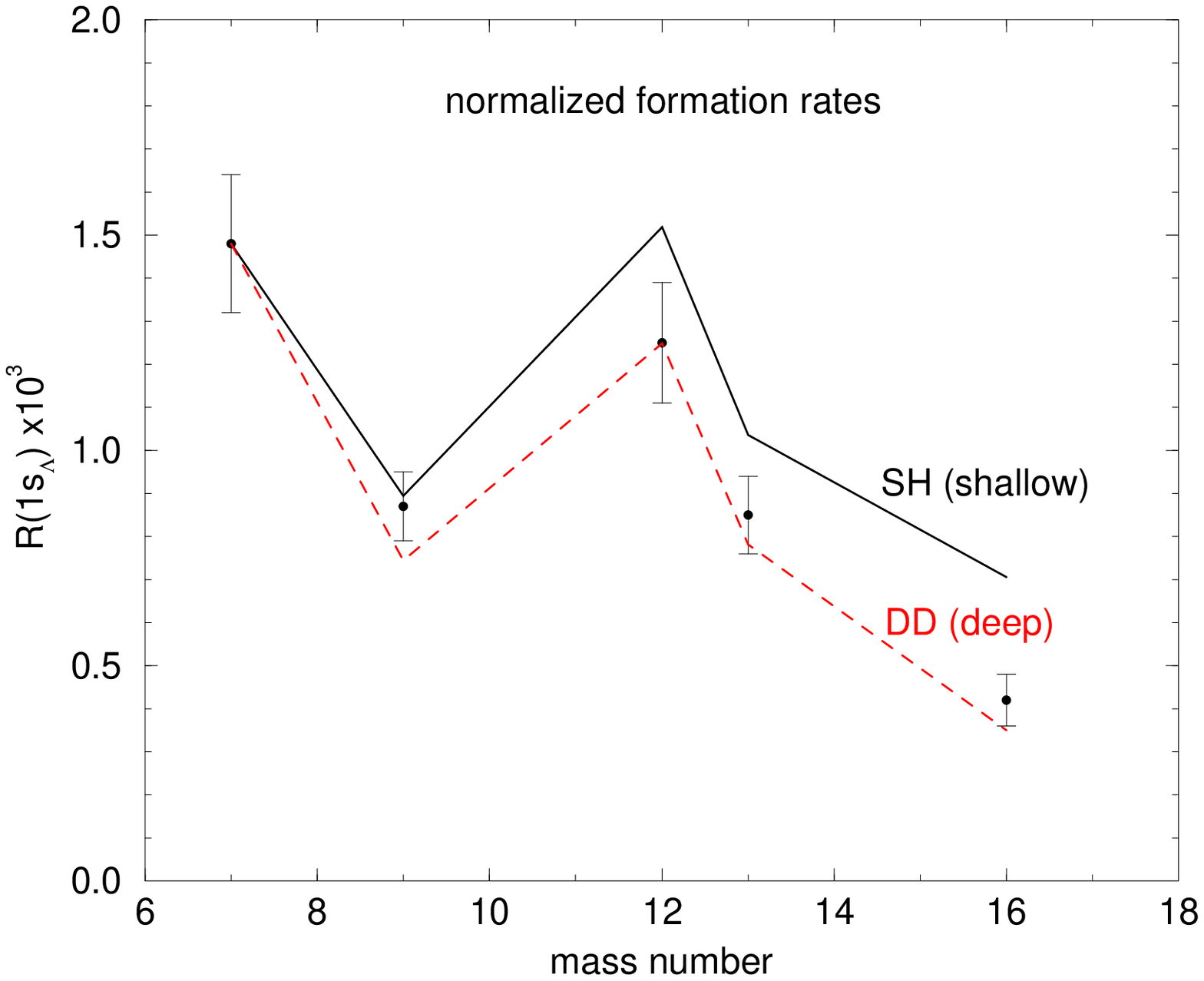} 
\end{center} 
\caption{Left: overlap of $K^-$-Ni atomic 4$f$ radial wavefunction $R$ 
squared with Ni matter density $\rho_{\rm m}$ in two models \cite{FG11}. 
Right: $1s_{\Lambda}$ formation rates per $K^-_{\rm stop}$ from $K^-$ capture 
at rest spectra taken by FINUDA \cite{FIN11} and as calculated \cite{CFGK11} 
-- normalized to the $_{\Lambda}^7$Li datum -- using in-medium density 
dependent $K^-n\to \pi^-\Lambda$ branching rates that relate subthreshold 
energies to densities similarly to Eq.~(\ref{eq:SC}).} 
\label{fig:gal4} 
\end{figure}

Several other fitted kaonic atom potentials are compared to the IHW-based 
(1$N$+2$N$) potential in Table~\ref{tab:DD}. The DD potential is a purely 
phenomenological potential of a form similar to Eq.~(\ref{eq:phen}) and 
offers a benchmark, with $\chi ^2$=103, for what may be viewed as the 
ultimate density dependent fit to 65 data points across the periodic table 
(it was denoted {\it nominal} in Ref.~\cite{FGB94}). The entry for the NLO30 
model is typical of results obtained in Refs.~\cite{CFGGM11,FG12}. All three 
displayed fits produce deeply attractive real potentials, with depth in the 
range 180--200 MeV at the center of Ni, and sizable absorptivities measured 
by imaginary potential depths in the range 70--80 MeV. The r.m.s radii of 
$V_{\rm R}$ are all smaller significantly than the point-proton distribution 
r.m.s. radius $r_p$=3.69~fm in Ni, reflecting the sizable contribution of 
the more compact Re~$V^{(2)}_{K^-}$, whereas the r.m.s radii of $V_{\rm I}$ 
are all slightly larger than $r_p$, reflecting the compensating effect of 
Im~$V^{(2)}_{K^-}$ on the rapidly decreasing with density Im~$V^{(1)}_{K^-}$. 
The very significant improvement of 30 units in $\chi^2$ values by going from 
NLO30 to IHW is due to species where strong interaction observables were 
measured for more than a single kaonic atom level. The width (or equivalently 
`yield') of the upper level is normally dominated by Im~$V^{(1)}_{K^-}$, 
whereas the width of the lower level is dominated by Im~$V^{(2)}_{K^-}$. 
{\bf Thus, more accurate determination of two level widths in the same kaonic 
atom are likely to pin down the density dependence of Im~$V^{(2)}_{K^-}$ as it 
evolves with density and overtakes Im~$V^{(1)}_{K^-}$.} The range of nuclear 
densities which prove to be effective for absorption from the lower level is 
exhibited on the l.h.s. of Fig.~\ref{fig:gal4} for Ni by plotting overlaps 
of the 4$f$ atomic radial wavefunction squared with the Ni matter density 
$\rho_{\rm m}$ for two choices of $V_{K^-}$ \cite{FG11}; 
see also Refs.~\cite{BF07,YH07}. The figure demonstrates that, 
whereas this overlap for the relatively shallow, density-independent 
$t\rho$ potential peaks at nuclear density of order 10\% of $\rho_0$, 
it peaks at about 60\% of $\rho_0$ for the deeper, density-dependent 
DD potential and has a secondary peak well inside the nucleus 
(indicating that a $K^-$ nuclear $\ell=3$ quasibound state exists). The DD 
potential, clearly, exhibits sensitivity to the interior of the nucleus 
whereas the $t\rho$ potential exhibits none. {\bf The superiority of Deep 
to Shallow $V_{K^-}$ can be checked by devising new measurements in a few 
carefully selected kaonic atoms \cite{elifried11}.} 

A reaction that could discriminate between deep and shallow $K^-$ nuclear 
potentials is the $K^-$ capture at rest formation of $\Lambda$ hypernuclear 
states localized in the nuclear interior. The formation rates are expected 
to be sensitive to the extent to which the relevant $K^-$ atomic wavefunctions 
penetrate into the nucleus. Spectra and formation rates of several $p$-shell 
hypernuclei were reported recently by the FINUDA experiment \cite{FIN11} and 
analyzed by Ciepl\'{y} et al. in Ref.~\cite{CFGK11}. A comparison between 
experiment and calculation is shown on the r.h.s. of Fig.~\ref{fig:gal4}. 
The calculations use in-medium subthreshold $K^-n\to \pi^-\Lambda$ branching 
rates where the density dependence is related to the subthreshold energy 
according to Eq.~(\ref{eq:SC}). {\bf This comparison favors deep potentials 
to shallow ones.} 

\begin{figure}[htbp] 
\begin{center} 
\includegraphics[width=0.48\columnwidth]{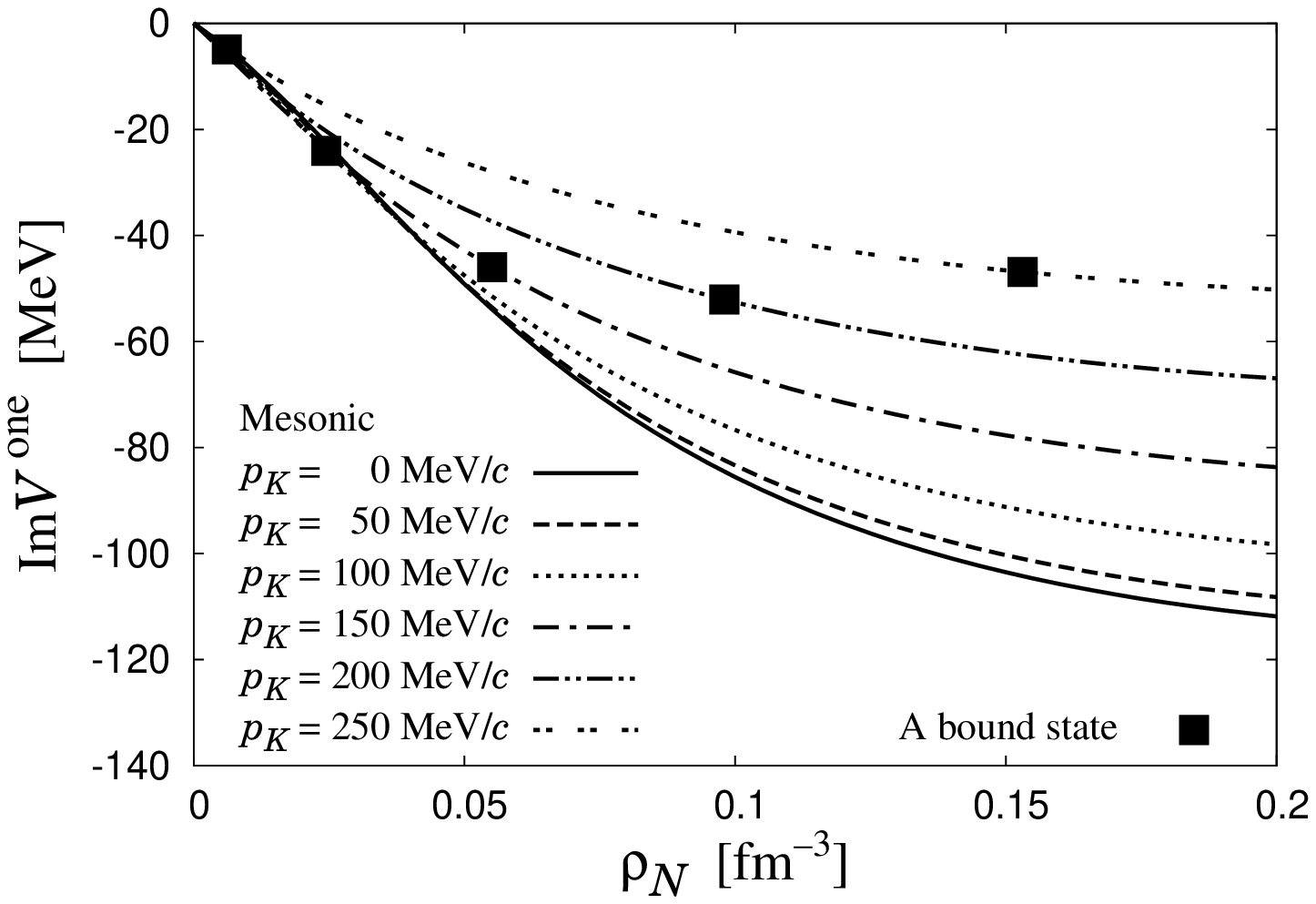} 
\includegraphics[width=0.48\columnwidth]{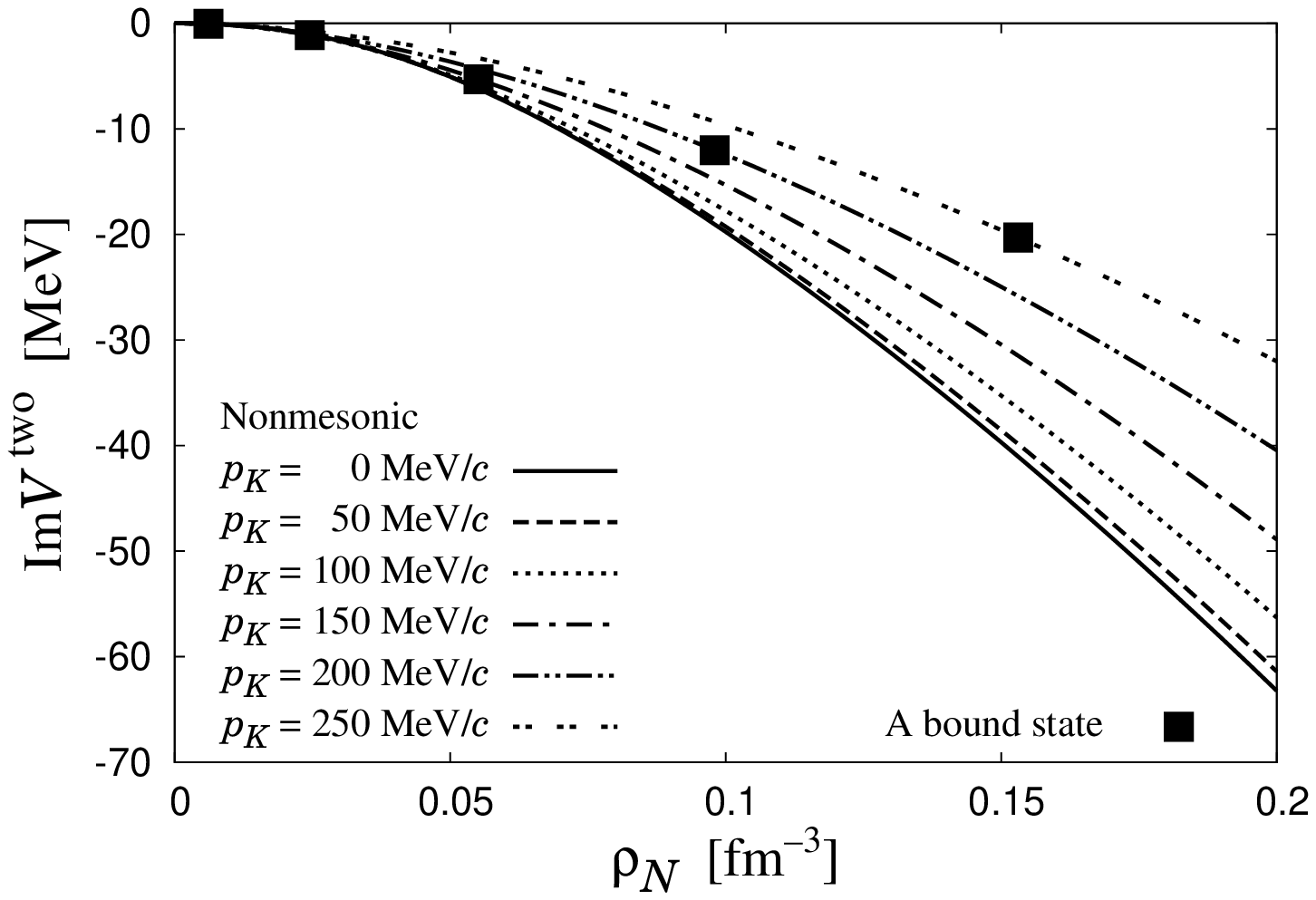} 
\end{center} 
\caption{Im~$V^{(1)}_{K^-}(\rho_N)$ (left) and Im~$V^{(2)}_{K^-}
(\rho_N)$ (right) evaluated in nuclear matter within a chiral unitary 
approach \cite{Sek12} for several values of the $K^-$ momentum $p_K$. 
The filled squares denote the density $\rho_N$ corresponding to $p_K$, 
see text.} 
\label{fig:gal5} 
\end{figure} 

Attempts to evaluate the absorptivity Im~$V^{(2)}_{K^-}$ in addition to 
Im~$V^{(1)}_{K^-}$ within a chiral unitary approach have been reported by 
Sekihara et al.~\cite{Sek12}. Their calculated nuclear-matter potentials, 
as a function of nuclear density $\rho_N$, are shown in Fig.~\ref{fig:gal5} 
for several values of the $K^-$ momentum $p_K$ between 0 to 250~MeV/c. 
The higher $p_K$ is, the deeper the $K^-N$ related amplitudes bite into 
the subthreshold region, which weakens both 1$N$ and 2$N$ absorptivities, 
in agreement with the lessons gained in Refs.~\cite{FG13,CFGGM11,FG12,GM12}. 
To establish correspondence with finite nuclei, these authors determined 
the local densities probed for the chosen values of $p_K$ by offsetting the 
$K^-$ kinetic energy $p_K^2(\rho_N)/2m_K$ against a prescribed potential 
$V_{K^-}(\rho_N)$=$-$70~MeV~$\times(\rho_N/\rho_0)$, resulting in values 
of density $\rho_N$ marked by filled squares in the figure. 
The finite-nucleus absorptivities are then obtained by connecting smoothly 
these filled squares. Comparing these calculated absorptivities at nuclear 
central density $\rho_0\approx 0.16\pm 0.01~{\rm fm}^{-3}$ with the IHW-based 
absorptivities shown for Ni in the lower r.h.s. of Fig.~\ref{fig:gal3}, 
one observes that Im~$V^{(1)}_{K^-}$ is overestimated by over a factor of 
two whereas Im~$V^{(2)}_{K^-}$ is underestimated by over a factor of three.
As a consequence, the ratio of Im~$V^{(2)}_{K^-}$/Im~$V^{(1)}_{K^-}$ plotted 
in Fig.~17 of Sekihara et al.~\cite{Sek12} underestimates significantly 
the relative strength of $K^-$ multinucleon absorption (occurring at $\rho 
\lesssim 0.5 \rho_0$) as compared to indications from old emulsion and 
bubble-chamber work \cite{VVW77}.

\section{Many-body kaonic quasibound states} 
\label{sec:many} 

\begin{figure}[htbp]
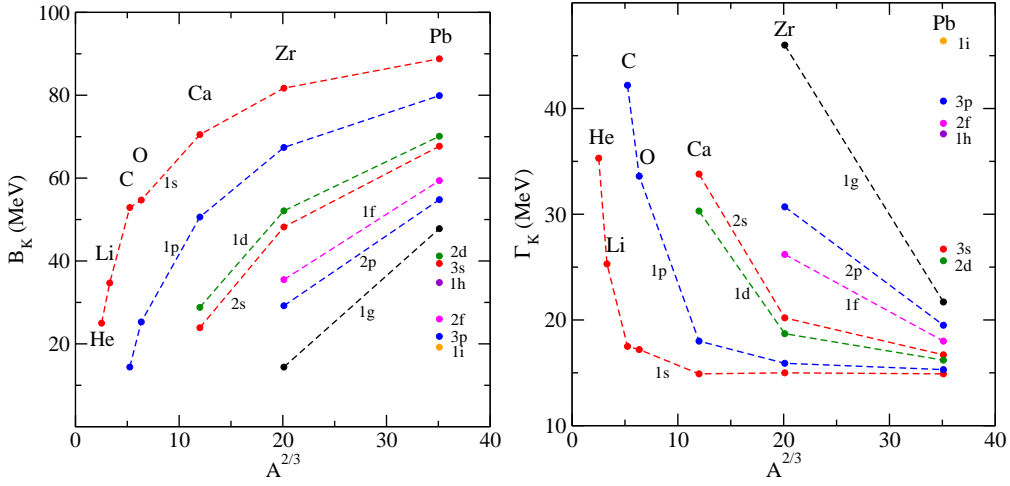
 
\begin{center} 
\includegraphics[width=0.48\columnwidth]{gm12_fig6a.eps} 
\includegraphics[width=0.48\columnwidth]{gm12_fig6b.eps} 
\end{center} 
\caption{Binding energies $B_K$ (left) and widths $\Gamma_K$ (right) of $K^-$ 
quasibound states in selected nuclei calculated self consistently by Gazda and 
Mare\v{s} \cite{GM12} using static RMF densities and NLO30 in-medium $K^-N$ 
subthreshold amplitudes, but without any $V^{(2)}_{K^-}$ contribution.} 
\label{fig:gal6} 
\end{figure} 

The state of the art in $K^-$ quasibound states calculations in nuclear 
systems beyond four-body systems has been reviewed recently, and also 
in these proceedings, by Ji\v{r}\'{i} Mare\v{s} \cite{GM12}. For the sake 
of completeness, I show in Fig.~\ref{fig:gal6} results borrowed from his 
presentation summarizing self consistent calculations of $K^-$ quasibound 
states in nuclei ranging from $^4$He to $^{208}$Pb. 
A prominent feature resulting from the imposition of self consistency on 
these calculations is the hierarchy of widths calculated within a given 
nucleus. With energy independent potentials one expects the width to be 
maximal for the lowest, most localized $1s_K$ states, and to monotonically 
decrease for excited states which are less localized within the core nucleus. 
The reverse is observed on the r.h.s. of the figure. This arises by requiring 
self consistency: the more excited a $K^-$ quasibound state is, the lower 
nuclear density it feels, and a smaller downward shift into subthreshold 
energies is implied by the $s_{\rho}$ dependence. Since Im~$V^{(1)}_{K^-}$ 
decreases strongly below threshold, this means that its contribution gets 
larger, the higher in energy the excited quasibound state is. As stressed by 
Gazda and Mare\v{s} \cite{GM12}, {\bf the large widths coupled with smaller 
excitation energy spacings should make it difficult to identify experimentally 
such $K^-$ quasibound states in a unique way. Additional width contributions 
from Im~$V^{(2)}_{K^-}$ will only compound this difficulty.}

\section{$\Sigma^{\ast}(1385)N$ vs. $\Lambda^{\ast}(1405)N$} 
\label{sec:piassist} 

The $I=1/2$, $J^{\pi}=0^-$ $K^-pp$ quasibound state discussed in 
Sect.~\ref{sec:few} represents a close realization of a $\Lambda^{\ast}N$ 
quasibound $^{1}S_0$ dibaryon \cite{UHO11}. Similarly, one could ask whether 
$\Sigma^{\ast}N$ dibaryon quasibound states are realized and, if so, for 
which quantum numbers. A close realization of a $\Sigma^{\ast}N$ quasibound 
$^{5}S_2$ dibaryon in terms of $I=3/2$, $J^{\pi}=2^+$ $\pi YN$ quasibound 
state $\cal Y$ was proposed a few years ago and established more solidly in 
the recent work of Ref.~\cite{GG13} within a relativistic three-body Faddeev 
calculation some 10--20 MeV below the $\pi\Sigma N$ threshold. The leading 
two-body attractive channels in this three-body system are the $p$-wave 
$\pi N$ and $\pi\Lambda-\pi\Sigma$ channels dominated by the $\Delta(1232)$ 
and $\Sigma(1385)$ resonances, respectively, and to a lesser extent the 
$^{3}S_1$ $YN$ $s$-wave channel. It was found that admixing a $\bar KNN$ 
channel to the $\pi YN$ three-body channels affects little this dibaryon 
quasibound state since for a $p$-wave $\bar K$ meson it requires a Pauli 
forbidden $I_{NN}=1$, $S_{NN}=1$ combination for the leading $NN$ $s$-wave 
configuration. 

{\bf This dibaryon candidate $\cal Y$ may be looked for in experiments similar 
to those being run at GSI \cite{Fabbietti12} and J-PARC \cite{NagaeE27} which 
aim at observing the associated production of $K^-pp$ and its subsequent decay 
to a $\Lambda p$ pair.} Thus, at GSI one could look for  
\begin{eqnarray} 
   p~ + ~p & ~\rightarrow ~ & {\cal Y}^{++} ~+~K^0 \nonumber  \\  
           &                & ~\hookrightarrow ~ \Sigma^+ ~+~ p \; ,
\label{eq:pptoY++} 
\end{eqnarray} 
where the decay ${\cal Y}^{++} \to \Sigma^+ p$ offers a unique decay channel, 
and at J-PARC look for 
\begin{eqnarray} 
\pi^{\pm} ~+~ d & ~\rightarrow ~ & {\cal Y}^{++/-} ~+~K^{0/+}  \nonumber  \\ 
 &  &  ~\hookrightarrow ~ \Sigma^{\pm} +p(n) \; , 
\label{eq:pi+dtoY++} 
\end{eqnarray} 
where again these decay channels offer unique channels distinctly from 
those looked for in $K^-pp$ searches.

\section*{Acknowledgments}

Special thanks are due to Nir Barnea, Ale\v{s} Ciepl\'{y}, Eli Friedman, 
Humberto Garcilazo, Daniel Gazda and Ji\v{r}\'{i} Mare\v{s}, with whom 
I have collaborated on topics surveyed here, and to Wolfram Weise for 
many ongoing stimulating discussions and kind hospitality at TUM in 
recent years.

\end{document}